# Fermionic Order by Disorder in a van der Waals Antiferromagnet

R. Okuma[1*], D. Ueta[1*], S. Kuniyoshi[1,2*], Y. Fujisawa[1], B. Smith[1], C. H. Hsu[1,3], Y. Inagaki[4], W. Si[4], T. Kawae[4], H. Lin[5], F. C. Chuang[3,6], T. Masuda[7], R. Kobayashi[2], Y. Okada[1]

mail: yoshinori.okada@oist.jp

[1]*Quantum Materials Science Unit, Okinawa Institute of Science and Technology (OIST), Okinawa 904-0495, Japan*

[2]*Faculty of Science, University of the Ryukyus, Nishihara, Okinawa 903-0213, Japan*

[3]*Department of Physics, National Sun Yat-sen University, Kaohsiung 80424, Taiwan*

[4]*Department of Applied Quantum Physics, Kyushu University, Fukuoka 819-0395, Japan*

[5]*Institute of Physics, Academia Sinica, Taipei, Taiwan*

[6]*Physics Division, the National Center for Theoretical Sciences, Hsinchu, 30013, Taiwan*

[7]*Institute for Solid State Physics (ISSP), The University of Tokyo, Kashiwa, Chiba 277-8581, Japan*

**Abstract**

$CeTe_3$ is a unique platform to investigate the itinerant magnetism in a van der Waals (vdW) coupled metal. Despite chemical pressure being a promising route to boost quantum fluctuation in this system, a systematic study on the chemical pressure effect on $Ce^{3+}(4f^1)$ states is absent. Here, we report on the successful growth of a series of Se doped single crystals of $CeTe_3$. We found a fluctuation driven exotic magnetic rotation from the usual easy-axis ordering to an unusual hard-axis ordering. Unlike in localized magnetic systems, near-critical magnetism can increase itinerancy hand-in-hand with enhancing fluctuation of magnetism. Thus, seemingly unstable hard-axis ordering emerges through kinetic energy gain, with the self-consistent observation of enhanced magnetic fluctuation (disorder). As far as we recognize, this order-by-disorder process in fermionic system is observed for the first time within vdW materials. Our finding opens a unique experimental platform for direct visualization of the rich quasiparticle Fermi surface deformation associated with the Fermionic order-by-disorder process. Also, the search for emergent exotic phases by further tuning of quantum fluctuation is suggested as a promising future challenge.

**Introduction**

In the search for interesting metallic states in quantum materials, controlling magnetic orientation, fluctuations, and their interaction with conduction electrons has been of utmost importance. Consequent exotic phases are rich, ranging from unconventional superconductivity[1,2,3,4] to giant magneto-transport effects, including metal-to-insulator transition [5], magnetically driven 2D confinement[6], and exotic Hall effects[7,8,9,10,11,12,13,14]. One of the most important recent challenges within this field is exploring the exotic magnetic metallic state in van del Waals (vdW) coupled materials[15,16,17,18,19,20,21]. A missing but crucial component in this challenge is establishing various approaches to controlling the interaction between magnetism and conduction and the mechanism by which they operate.

$CeTe_3$ provides a unique platform to investigate the itinerant magnetism in a van der Waals (vdW) coupled metal[22,23,24]. The crystal structure of $CeTe_3$ features a square net of $Te^{0.5-}$ and rock-salt type layer of $Ce^{3+}$ and $Te^{2-}$ (**Fig. 1a**), combined to realize a highly two-dimensional vdW motif[21, 25]. The conduction band mainly originates from the $Te^{0.5-}$ square net; $Ce^{3+}$ is responsible for magnetism. Strong nesting of the Fermi surface derived from the square net Te sheet causes a 1D-like charge density wave (CDW) transition below 400 K[26,27,28,29,30], whose temperature scale is well separated from the much lower magnetic transition temperature. Based on previous studies[23,31], the magnetic moment of $CeTe_3$ can be described as a pseudo spin 1/2 system with strongly easy-plane (XY) character, and long range antiferromagnetic order along the easy-plane is observed below 1.3 K.

Recently, $CeSeTe_2$ has been identified as a compound closely related to $CeTe_3$. In $CeSeTe_2$, $Se^{2-}$ selectively substitutes into $Te^{2-}$ sites in $CeTe_3$[32,33]. Consistent with the chemical and structural similarity of $CeSeTe_2$ to $CeTe_3$ (**Fig. 1a**), a CDW transition and strongly easy-plane like magnetism in the paramagnetic regime are observed in both materials[31]. In $CeSeTe_2$ however, the reported magnetism in the ordered state is consistent with a collinear structure pointing in the out-of-plane direction[31], which is strongly unfavorable in the paramagnetic state. Thus, by defining the magnetic hard and easy axis/plane of the system based on its paramagnetic state, surprising *hard-axis ordering* is suggested to emerge in $CeSeTe_2$ below the antiferromagnetic phase transition. Despite this unusual direction of magnetic moment in the ordered state, no detailed measurements have been performed so far to elucidate the underlying mechanism.

To understand the origin of this peculiar hard-axis ordering, here we studied the effect of Se doping systematically through magnetization and heat capacity measurement of $Ce(Se_xTe_{1-x})Te_2$. We confirmed that addition of Se applies chemical pressure to $Ce^{3+}$, which results in exotic reorientation of the magnetic moment through enhanced quantum fluctuation. In this report, we present that the presence of hard-axis ordering is representative of the fermionic order by disorder process, in which a seemingly unstable direction is stabilized through itinerant energy gain.

**Results**

**Crystal growth of Ce(Se$_x$Te$_{1-x}$)Te$_2$.** Single crystals of Ce(Se$_x$Te$_{1-x}$)Te$_2$ were grown via the flux method (See method for detail). This formula represents that Se$^{2-}$ selectively replaces Te$^{2-}$ in the magnetic blocking layer (**Fig. 1a**). Thus, Se$^{2-}$ doping connects the two isostructural systems Ce$^{3+}$(Te$^{2-}$)(Te$^{0.5-}$)$_2$ and Ce$^{3+}$(Se$^{2-}$)(Te$^{0.5-}$)$_2$. Since the ionic radii of Se$^{2-}$ and Te$^{2-}$ are 1.98 Å and 2.21 Å, respectively[34], systematic control of chemical pressure becomes possible in Ce(Se$_x$Te$_{1-x}$)Te$_2$ through variation of the doping ratio. The single crystals of Ce(Se$_x$Te$_{1-x}$)Te$_2$ present a plate-like morphology with shiny surfaces, reflecting the vdW coupling nature in our system (**Fig. 1a**). **Fig. 1b** shows the doping evolution of the X-ray diffraction (XRD) pattern around the (080) peak. **Fig. 1c** shows the relationship of the lattice constant along the out-of-plane b axis and the Se/Te ratio x as determined by XRD and energy dispersive X-ray spectrometry (EDX), respectively. As predicted by Vegard's law, the monotonic linear behavior of the shrinkage of lattice constant with doping suggests the systematic replacement of smaller ionic radii Se$^{2-}$ into the larger ionic radii Te$^{2-}$ sites. Similar monotonic shrinkage of the out of plane lattice constant by doping with elements with smaller ionic radii is also seen in another vdW misfit compounds such as Bi-based cuprates[34]. In this study, using the obtained linear function, we calculated *x* from the out-of-plane lattice constant of the crystal flakes used for heat capacity and magnetic measurements.

**Magnetic order in CeSeTe$_2$ and CeTe$_3$.** We first present a significant difference in the magnetism and heat capacity behavior of CeTe$_3$ (**Fig. 2a**) and CeSeTe$_2$ (**Fig. 2b**), despite the common easy-plane feature in the paramagnetic regime. In **Fig. 2a and b**, magnetic susceptibility $\chi$(T) and heat capacity $C$(T) are shown on the left and right axes, respectively. For the parent compound CeTe$_3$, we see two characteristic antiferromagnetism related transition temperatures $T_{N1}$ and $T_{N2}$. Only a broad peak is observed in $C$(T) around $T_{N1}$ while around $T_{N2}$ a sharp peak in $C$(T) accompanies stronger suppression of $\chi$(T) along the in-plane (blue and red curves) than the out-of-plane (greed curve) direction. On the other hand, the doped system CeSeTe$_2$ features a single antiferromagnetic transition with a sharp peak at T$_{N3}$ in $C$(T) that accompanies stronger suppression of $\chi$(T) along the out-of-plane (green curve) than the in-plane (blue and red curves) directions. Therefore, at the lowest temperature, the Néel vector points along the easy plane in CeTe$_3$, whereas it points along the hard axis in CeSeTe$_2$. This striking contrast is also evident from the suppression of susceptibility and spin flip and flop transitions (**Fig. 2c and 2d**) since they should appear along the Néel vector. **Fig. 2e** is the experimentally determined magnetic hard axis and easy plane together with crystal axes. These directions are the same are the same for all samples shown in this study.

**Quasi two-dimensional magnetism of CeSeTe$_2$ and CeTe$_3$.** The ordering phenomena in Ce(Se$_x$Te$_{1-x}$)Te$_2$ based on localized spin picture are briefly discussed before we emphasize the importance of itinerancy. The magnetism of CeTe$_3$ and CeSeTe$_2$ is supposed to originate from highly two-dimensional interactions with XY-like and Ising-like anisotropy, respectively. In CeTe$_3$, the successive phase transitions associated with $T_{N1}$ and $T_{N2}$ are reminiscent of the 2D XY model perturbed by weak in-plane anisotropy[35]. While two-dimensional XY-like interactions can support only quasi long-range order at a finite temperature, genuine long-range order can appear by nearly four-fold symmetric anisotropies at a lower temperature. This picture is presumably captured by the broad and sharp peaks $T_{N1}$ and $T_{N2}$, respectively. In contrast, Ising-type long-range-order in CeSeTe$_2$ can appear by breaking discrete Ising symmetry; here in-plane anisotropy plays no role in the magnetic ordering and a single transition has been observed. The abrupt suppression of $\chi(T)$ also reflects on the stability of the Ising-like collinear order. It should be noted that the observed collinear magnetism suggests that the Dzyaloshinski-Moriya (DM) interaction, which favors noncollinear spin arrangement regardless of anisotropy of the g factor, should play a minor role in the emergence of hard axis magnetic order. At this stage, it remains unclear why the effective interaction becomes Ising-like by doping even though the paramagnetic susceptibility remains XY-like. As we discuss hereafter, beyond the localized magnetic picture, kinetic energy gain plays a crucial role in explaining an essential part of our experimental observation.

**Magnetic fluctuation of Ce(Se$_x$Te$_{1-x}$)Te$_2$.** If the ordering direction of CeSeTe$_2$ is actually unfavorable, we should observe self-consistent consequences of enlarged quantum fluctuation. To check this, we investigated the systematic doping dependence of the magnetic fluctuation of Ce(Se$_x$Te$_{1-x}$)Te$_2$. 4f-electron derived entropy $S(T)$ is estimated and plotted in **Fig. 3a-f** (right axis) based on the formula $S(T) = \int_0^T \frac{C_f}{T} dT$. Here, $C_i(T)$ is the heat capacity from the 4f electron contribution (See method for detail). We further calculate magnetic entropy $S_m$ for $S(T_{N1})$ and $S(T_{N3})$ as x<0.54 and x>0.54, respectively. Based on a fully localized picture, $S_m$ should asymptotically approach $R\ln2$ (~5.76 JK$^{-1}$mol$^{-1}$). This value is simply derived from the ground state doublet of Ce$^{3+}$ ions under the crystalline electric field. Thus, 1-$S_m$/$R\ln2$ can be taken as the measure of the strength of magnetic fluctuation for a given magnetic order.

**Evolution of magnetic fluctuation by doping.** By constructing a phase diagram, a correlation exists between enhanced magnetic fluctuation and emergent hard-axis moment. **Fig. 4a** represents the phase diagram with three characteristic temperatures (left axis) together with 1-$S_m/R$ln2 (right axis). With increasing doping x, $T_{N1}$ and $T_{N2}$ are systematically decreased for x<0.54. Above x=0.54, we see only one transition temperature $T_{N3}$. Therefore, combining magnetic measurements for CeTe$_3$ (**Fig. 2a**) and CeSeTe$_2$ (**Fig. 2b**), a first-order magnetic transition is expected to exist near x=0.54, across which the direction of magnetic moment abruptly changes from the easy plane to the hard axis. The enhancement of 1-$S_m/R$ln2 towards the first-order transition point is partly due to the chemical disorder effect since chemical entropy naturally reaches a maximum at x=0.5 in our alloyed system. As a consequence, chemical disorder may contribute to a slight increase in 1-$S_m/R$ln2 with doping levels above x = 0.54 (**Fig. 4a**). However, we emphasize that the values 1-$S_m/R$ln2 for two nearly stoichiometric samples with x=0 and x=0.96 are 0.25 and 0.36, respectively. Based on these values, the magnetic fluctuation of stoichiometric CeSeTe$_2$ is expected to be ~1.4 times larger than that of the parent compound CeTe$_3$. Therefore, the key feature to be captured in the phase diagram is an emergent hard-axis moment with a monotonic enhancement of magnetic fluctuation.

**Discussion**

We propose the possible mechanism leading to emergent hard axis ordering in terms of fermionic order by disorder in itinerant systems[36,37]. Our observation is that the change in chemical pressure enhanced the magnetic quantum fluctuation and altered the direction of Néel vector to the apparent "hard axis". This implies that magnetic quantum fluctuation itself drives the hard-axis ordering as in the case of "order by disorder" in frustrated magnets[38]. The relevant quantum fluctuation is unlikely to be Kondo screening in our case because a similar magnitude of saturated magnetic moment along the easy-axis with doping was observed experimentally (**Fig. 2c** and **d**). Instead, we propose that the enhanced magnetic fluctuation is a spin-wave excitation, which is the primary fluctuation from the magnetically ordered state. An important point to note is that the spin-wave excitation is a transverse fluctuation. Since the transverse direction from the hard axis is on the easy plane, hard-axis order can fluctuate more easily than easy-plane order. We surmise that the hard-axis order becomes favorable when chemical pressure further enhances quantum fluctuation of localized spin through Kondo coupling.

Compared to localized frustrated magnets, fermionic order by disorder requires kinetic energy gain of electrons in addition to transverse spin fluctuation (**Fig. 4b**). Such physics is directly captured by investigating electronic quasiparticle band deformation to lower the total energy of the system. This concept is consistent with the Fermionic order-by-disorder process, as predicted theoretically[36,37]. Regardless of theoretical proposals, however, experimental direct visualization of quasi-particle band deformation has been missing due to the lack of proper materials to investigate with high resolution. Compared with materials in existing studies such as YbRh$_2$Si$_2$, YbNi$_4$P$_2$, and Ce$T_2$Al$_{10}$ ($T$ = Ru, Os)[39,40,41,42,43,44,45] the intrinsically cleavable nature and simple composite structure of our vdW material offers an excellent platform for unveiling rich quasiparticle band deformation using surface sensitive spectroscopies with changing external magnetic field, its orientation and temperature. In addition, inherent applicability of advanced exfoliation techniques [46] will further enrich future challenges in searching for novel quantum phenomena.

In summary, we show magnetic fluctuation driven unusual magnetic rotation from the usual easy plane to unusual hard axis moment across the magnetic phase transition. As far as we recognize, this is the first report of the order-by-disorder process in fermionic system in a vdW coupled antiferromagnet, which opens various unique research directions for the future.

## Methods

**Synthesis of monocrystalline Ce(Se$_x$Te$_{1-x}$)Te$_2$**

Single crystals of Ce(Se$_x$Te$_{1-x}$)Te$_2$ are grown using self-flux method. The mixture of the elemental powder of cerium (99.9%), selenium (99.999%), and tellurium (99.999%) in a molar ratio of 1 : 2x/3 : 40-2x/3 was placed inside the alumina crucible and sealed in an evacuated quartz tube. The ampoule was heated to 900°C at a rate of 75°C/hour, kept at 900°C for 24 hours, and then slowly cooled down to 550°C at a rate of 2°C/hour followed by centrifugation to remove crystals from the melt.

**Magnetization and heat capacity measurement**

DC magnetization and heat capacity were measured using a Magnetic Property Measurement System (Quantum Design) with a 3He insert [47] and a Physical Property Measurement System PPMS (Quantum Design) with a 3He insert, respectively.

**Estimation of magnetic heat capacity**

We extracted the magnetic fluctuation strength from heat capacity $C(T)$ (**Fig. 3a-f left axis**) based on the method described as follows. To obtain the 4f electron contribution $C_f(T)$, first we subtract $C(T)$ of LaTe$_3$ from that of Ce(Se$_x$Te$_{1-x}$)Te$_2$. This subtraction relies on the assumption that phonon and electronic contributions not arising from the 4f electrons are the same between LaTe$_3$ and Ce(Se$_x$Te$_{1-x}$)Te$_2$ (see supplemental information). From comparison, the specific heat is dominated by 4f electron derived signal, and phonon subtraction is a minor correction within the main purpose of this study. Using the obtained $C_f(T)$, Here, we set $C_f(0)=0$ to calculate $S(T)$. This assumption is physically not exactly accurate since heat capacity is finite due to itinerant 4f-electron nature; It does not however have a large impact within the purpose of extracting magnetic entropy since electronic contribution relative to magnetic contribution is much lower.

**Data availability**

The datasets collected and analysis performed by this study are available from the corresponding author upon reasonable request.

**Acknowledgements**

The cell structure was visualized using VESTA[48]. Heat capacity measurements were carried out under the Visiting Researcher's Program of Institute for Solid State Physics, The University of Tokyo. We thank S. Asai and S. Hasegawa for helping with heat capacity measurements.

**Author contribution**

Y. O. and R. K. conceived this project. D. U, S. K., Y. F., R. K., Y. O., B. S., Y. I., W. S., T. M., and T. K. corrected experimental data. C. H., H. L., and F. C., theoretically considered electronic structure. R. O. and Y. O. interpret the data. Y. O., R. O., and B. S. wrote the manuscript. We all discussed.

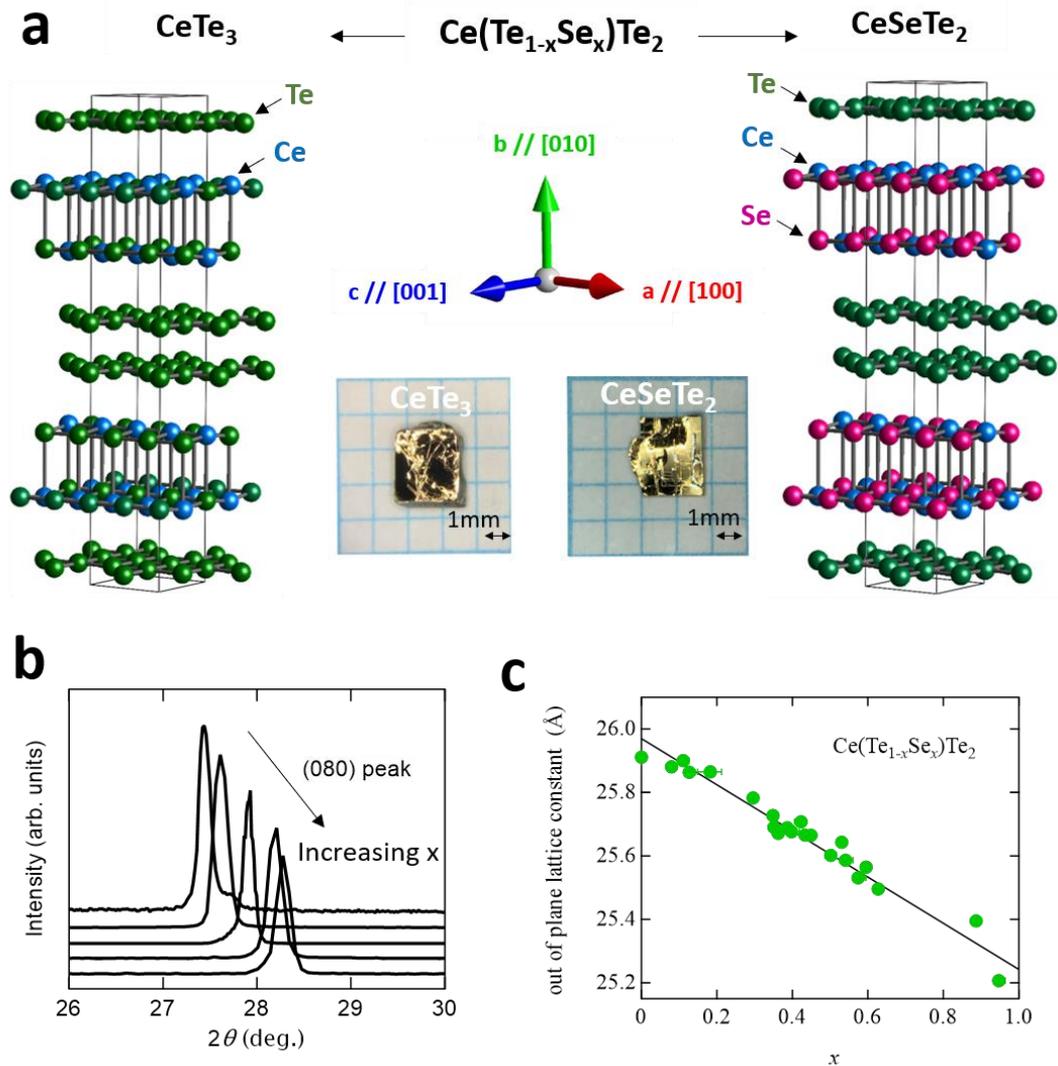

**Fig. 1 Systematic change of chemical pressure by iso-valent elemental substitution in a vdW coupled material Ce(Se$_x$Te$_{1-x}$)Te$_2$.** **(a)** Crystal structures of CeTe$_3$ (left) and CeSeTe$_2$ (right). Substituted Se atoms enter the magnetic blocking layer selectively. The typical picture of single crystals and the definition of crystallographic directions are also shown. **(b)** Doping dependence of characteristic X-ray diffraction (XRD) patterns near the (0 8 0) peak for Ce(Se$_x$Te$_{1-x}$)Te$_2$. **(c)** The out of plane lattice constant *b* as a function of doping x determined from energy dispersive X-ray spectrometry (EDX). This relation was obtained by performing both XRD and EDX on individual crystal flakes.

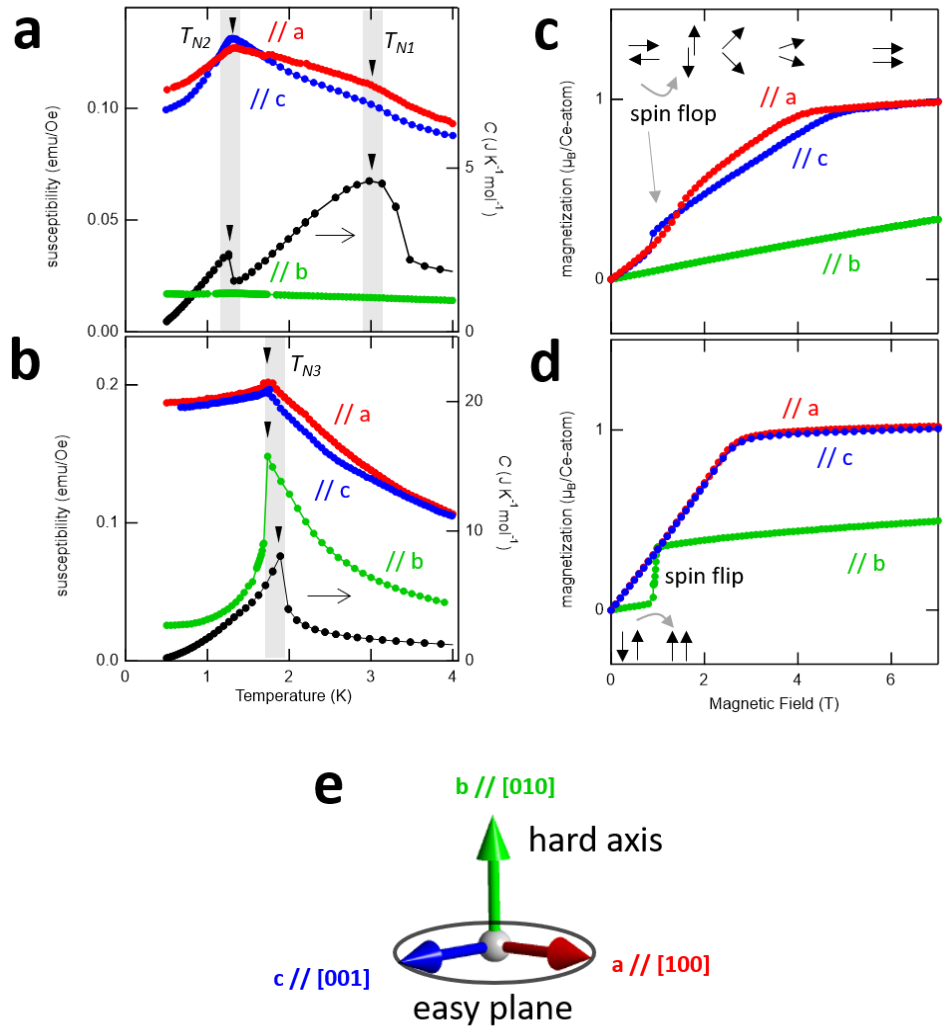

**Fig. 2 Magnetism and heat capacity characterization for CeTe$_3$ and CeSeTe$_2$ samples. (a,b)** Temperature dependence of the magnetic susceptibility (right axis) for (a) CeTe$_3$ and (b) CeSeTe$_2$ with applying external field $H$ = 0.1 T. On the right axis, temperature dependence of the heat capacity $C(T)$ is also shown. In **(a)** and **(b)**, there is slight difference in transition temperature since different samples were used for magnetic and heat capacity measurements. **(c, d)** Magnetic field dependence of the magnetization for (c) CeTe$_3$ and (d) CeSeTe$_2$ at $T$ = 0.5 K. Cartoons for the magnetization process with spin-flop along easy plane and spin-flip along hard axis are shown in (c) and (d), respectively. The red, blue, and green data shown in **(a)**-**(d)** are obtained for $H$ // a, $H$ // c, and $H$ // b, respectively. **(e)** The experimentally determined magnetic hard axis and easy plane together with crystal axes. These crystallographic and magnetic directions are the same for all samples shown in this study.

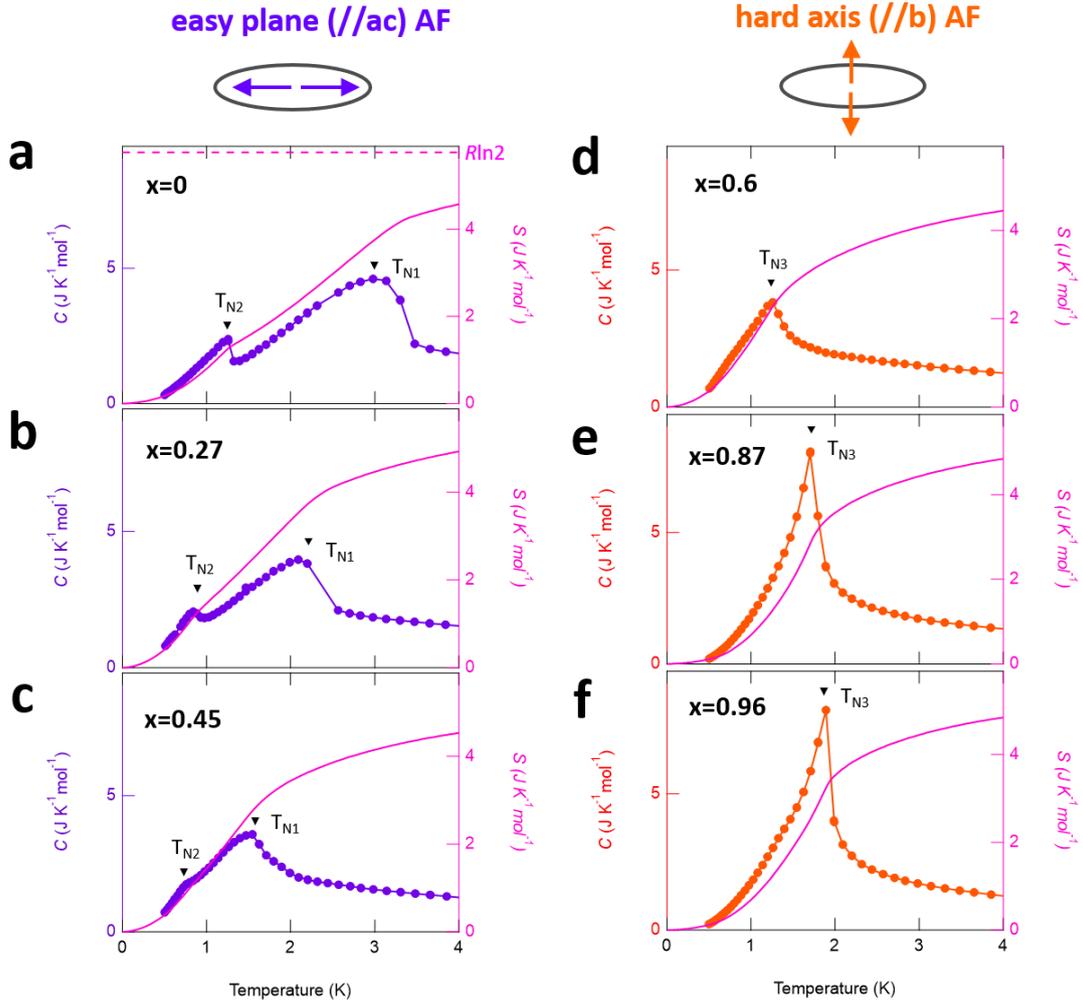

**Figure 3 Doping and temperature dependence of heat capacity $C(T)$ of Ce(Se$_x$Te$_{1-x}$)Te$_2$. (a-f)** Temperature dependence of the heat capacity (circles; left axis) and magnetic entropy (solid line; right axis) of (**a**) x=0, (**b**) x=0.27, (**c**) x=0.45, (**d**) x=0.60, (**e**) x=0.87, (**f**) x=0.96. The purple color represents data obtained from samples with easy plane antiferromagnetism (AF), and the orange represents those with hard axis AF. In (a), the value $R\ln 2$ (~5.76 JK$^{-1}$mol$^{-1}$) is shown with a broken line. This value is the calculated magnetic entropy from the ground state doublet of Ce$^{3+}$ ions under the crystalline electric field.

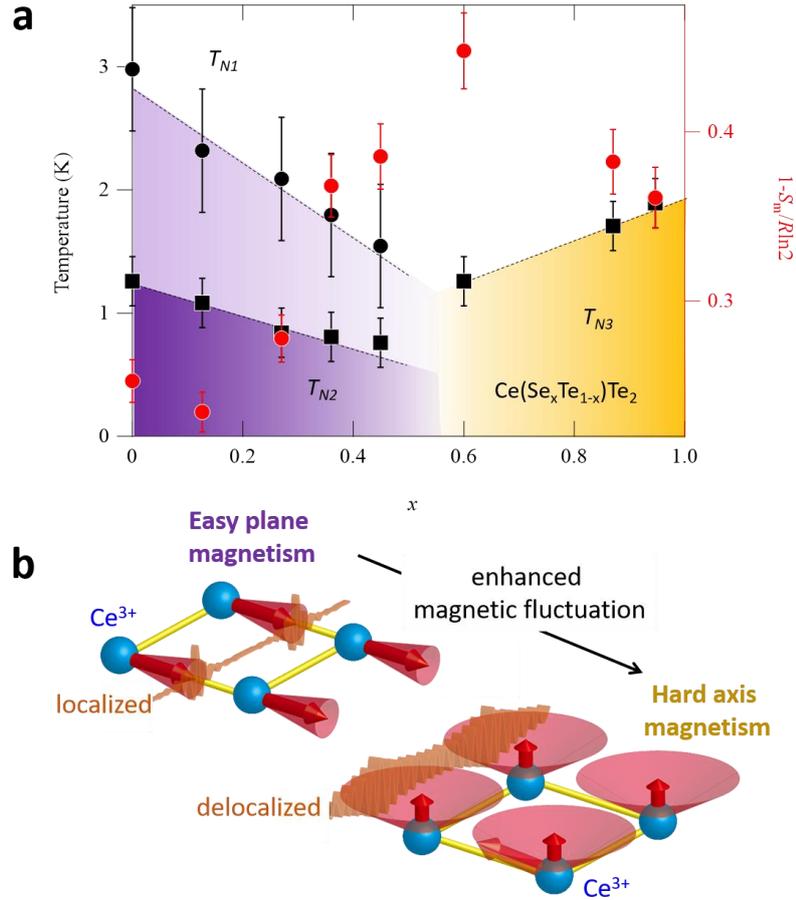

**Figure 4 The phase diagram representing magnetic rotation associated with enhanced quantum fluctuation.** (a) Temperature-doping phase diagram of $Ce(Se_xTe_{1-x})Te_2$ for three successive magnetic transition temperatures ($T_{N1}$, $T_{N2}$, and $T_{N3}$) and magnetic entropies $1-S_m/R\ln2$. The data used in this phase diagram are from specific heat measurements. (b) Schematic drawing of Fermionic order by disorder. The magnetic moment lies in the easy axis (plane) and is reduced with enhanced quantum fluctuation for $x < 0.54$. Whereas with enhanced fluctuation the magnetic moment moves to lie along the hard axis ($x > 0.54$) and expresses enhanced precession as for an Ising-like moment. The kinetic energy gain with enhanced magnetic fluctuation is represented as a change from localized wave packet (left) to delocalized wave packet (right). As a detailed spin structure for antiferromagnetism within this compound is not totally clear, ferromagnetically aligned spins within single Ce square lattice sheet along ac plane (see Fig. 1a and Fig. 2a) are drawn for clarity.